# Fast and Facile Synthesis Route to Epitaxial Oxide Membrane Using a Sacrificial Layer

Shivasheesh Varshney[1,*], Sooho Choo[1], Liam Thompson[2], Zhifei Yang[1,2], Jay Shah[1], Jiaxuan Wen[3], Steven J. Koester[3], K. Andre Mkhoyan[1], Alexander McLeod[2], and Bharat Jalan[1,*]

[1]Department of Chemical Engineering and Materials Science, University of Minnesota, Twin Cities, Minnesota, 55455, USA

[2]School of Physics and Astronomy, University of Minnesota, Twin Cities, Minnesota, 55455, USA

[3] Department of Electrical and Computer Engineering, University of Minnesota, USA

[*]Corresponding authors:

Bharat Jalan, Email: bjalan@umn.edu

Shivasheesh Varshney, Email: varsh022@umn.edu

**Keywords:** $SrTiO_3$ thin film | molecular beam epitaxy | membrane | synthesis | heterogeneous integration

# ABSTRACT

The advancement in thin-film exfoliation for synthesizing oxide membranes has opened up new possibilities for creating artificially-assembled heterostructures with structurally and chemically incompatible materials. The sacrificial layer method is a promising approach to exfoliate as-grown films from a compatible material system, allowing their integration with dissimilar materials. Nonetheless, the conventional sacrificial layers often possess intricate stoichiometry, thereby constraining their practicality and adaptability, particularly when considering techniques like Molecular Beam Epitaxy (MBE). This is where easy-to-grow binary alkaline earth metal oxides with a rock salt crystal structure are useful. These oxides, which include (Mg, Ca, Sr, Ba)O, can be used as a sacrificial layer covering a much broader range of lattice parameters compared to conventional sacrificial layers and are easily dissolvable in deionized water. In this study, we show the epitaxial growth of single-crystalline perovskite $SrTiO_3$ (STO) on sacrificial layers consisting of crystalline SrO, BaO, and $Ba_{1-x}Ca_xO$ films, employing a hybrid MBE method. Our results highlight the rapid ($\leq$ 5 minutes) dissolution of the sacrificial layer when immersed in deionized water, facilitating the fabrication of millimeter-sized STO membranes. Using high-resolution x-ray diffraction, atomic-force microscopy, scanning transmission electron microscopy, impedance spectroscopy, and scattering-type near-field optical microscopy (SNOM), we demonstrate epitaxial STO membranes with bulk-like intrinsic dielectric properties. The employment of alkaline earth metal oxides as sacrificial layers is likely to simplify membrane synthesis, particularly with MBE, thus expanding research possibilities.

**INTRODUCTION**

Modern electronic devices demand the integration of functional materials with diverse functionalities. Traditionally, the functional materials used in devices are grown on commercially available substrates using thin-film deposition techniques such as sputtering, Pulsed Laser Deposition (PLD), or MBE. The use of a rigid substrate clamps the thin film, imposing limitations such as reduced flexibility and bendability. Limitations in the choice of available substrates further restrict the degree of strain engineering. Additionally, the substrate clamping effect makes the creation of 3D architectures of such functional materials less feasible [1–5], raising the question of how to overcome these limitations. One way to address these limitations is by separating the film from the substrate. After exfoliation, the thin film membrane can be transferred on a desired host substrate, allowing integration with a number of structurally and chemically incompatible substrates, providing opportunities to develop symmetry-mismatched, non-equilibrium, and artificially-assembled heterostructures. These structures can further lead to the creation of moiré patterns and origami-inspired structures [6–9]. Using a flexible or stretchable substrate, an extreme strain (beyond a few percent in epitaxial systems) can further be achieved, leading to new functionalities, which otherwise is not possible in thin films clamped to a substrate [10–12].

A traditional method to create a freestanding membrane is via wet/dry etching of the crystalline substrate in acidic solutions or using lasers. This is a destructive approach that can potentially damage the membrane [13–15]. An alternative approach includes the remote epitaxy, where a thin van der Waals (vdW) layer such as graphene is inserted between the film and the substrate [7]. Epitaxial growth therefore requires growth conditions that are nondestructive to the vdW layer. Despite its successful application in creating membranes of diverse classes of materials, the underlying mechanism remains unclear [16,17]. Another technique uses an epitaxially-grown sacrificial layer in which a selectively etchable layer is inserted between the functional film and the substrate [3,18,19]. The sacrificial layer can be dissolved in water (or mild acidic solutions), and the developed membrane is transferred onto a desired host substrate. This approach promotes the reusability of the substrate after exfoliation of the functional film. It also has the advantage of growing the epitaxial sacrificial layer inside the same growth chamber with regulated growth conditions, atomic-level precision, and controlled interfaces [20].

When considering a material for use as an epitaxially-grown sacrificial layer, there are several criteria to take into account. These criteria may vary depending on the specific application, but some common factors include: (i) *Compatibility*: epitaxial growth of the single-crystalline functional film should be possible on top of the sacrificial layer without interfacial reaction/diffusion, (ii) *Selectivity*: the sacrificial layer should be etchable (preferably in non-acidic solutions) without damaging the functional film, (iii) *Stability*: the sacrificial layer should be thermally stable at the growth temperature which is typically > 800 °C for epitaxial oxide films, (iv) *Processability*: the sacrificial layer should be easily processable and reproducible using the existing equipment and processes and finally (v) *Tunability*: the sacrificial layer should have tunable lattice parameters for coherent, epitaxial growth of functional crystalline films [19–21]. Furthermore, it is essential to exercise caution to prevent potential harm to the functional film due to the deterioration of the sacrificial layer when exposed to air.

Various sacrificial layers have been developed [1,2,10,11,22]. For instance, $Sr_3Al_2O_6$ and its alloys $Ca_3Al_2O_6$ and $Ba_3Al_2O_6$ satisfy all the criteria and have been widely used as a sacrificial layer [19,23]. However, etching these sacrificial layers is time consuming, which can potentially affect their use in applications where long-exposure to water or moisture can degrade the target functional film. Additional sacrificial layers are $La_{0.7}Sr_{0.3}MnO_3$ [1,3], $SrVO_3$ [24], $YBa_2Cu_3O_7$ [2], $SrCoO_{2.5}$ [21], and SrRuO$_3$ [25]. These sacrificial layers have a complex stoichiometry optimization process, and the lattice parameter range accessed by them is limited to typical perovskite oxides (Figure 1). Specific to MBE, these sacrificial layers also contain low-vapor pressure elements (Co, Cu, Ru), magnetic elements (Mn, V, Co, and Ru), or deep acceptor-like impurity elements (Al), which possess synthesis challenges [26,27]. The use of these elements can further complicate the growth of epitaxial films due to the potential for cross contamination, particularly when considering techniques like MBE. To put this into context, we show, in Figure 1a, the number of publications on membrane synthesis using an epitaxial sacrificial layer since its first report in 2016. While there have been more than 70 publications on the growth of complex sacrificial layers using PLD, there are only limited publications using other growth approaches like chemical solution deposition (only 3)[28–30], and MBE (only 7)[10,31–36]. This shows PLD has been most successful in creating epitaxial membranes using a sacrificial layer. In part, this can be attributed to the difficulty associated with MBE for growing complex sacrificial layer. Simpler oxides such as BaO have also been employed as a sacrificial layer using PLD [37]. Nevertheless, there is no work concerning the utilization of binary oxides as sacrificial layers in MBE. Conversely, some efforts have been made to grow epitaxial complex oxides on binary oxides (*not* as a sacrificial layer), but these attempts have been largely unsuccessful [38–42] (see Figure S1).

The binary alkaline earth metal oxides (*A*O, *A*= Mg, Ca, Sr, and Ba) have great potential as a sacrificial layer in epitaxial membrane synthesis as they can also meet all the above criteria. They are easy to grow in MBE since the alkaline earth metals can be easily sublimated (300-500 °C) and readily oxidized. The binary oxides can also form a solid solution owning to the structural similarity (rock salt structure, Figure 1b), enabling tuning of lattice parameters over a wide range from 2.98 Å to 3.91 Å and 4.21 Å to 5.12 Å – a range currently unreachable using existing sacrificial layers and commercially-available substrates.

Figure 1c shows the lattice parameters of select functional oxides, commercially available substrates, and sacrificial layers using complex stoichiometry and binary alkaline earth oxides systems (Mg, Ca, Sr, Ba)O. The complex sacrificial layers cover the typical range of lattice parameters of perovskite oxides 3.81 Å to 4.12 Å. Combining these sacrificial layers with the binary system can provide a larger range of lattice parameters, providing the possibility of creating a membrane of not only lattice-matched functional oxides but also other functional materials beyond oxides[37,43–45]. For instance, (Mg,Ca)O system can potentially be used to make membranes of full Heusler ($X_2YZ$), half Heusler alloys (XYZ), III-V semiconductors, transition metal aluminides (TM-III), and rare-earth monopnictide, which have lattice parameters in the range 4.5 Å to 6.5 Å, twice the value of lattice parameter of (Mg,Ca)O system [46,47]. Future studies should investigate this possibility.

In this study, we demonstrate easy-to-synthesize sacrificial layers – SrO, BaO, and $Ba_{1-x}Ca_xO$ with a simple stoichiometry followed by the synthesis of high-quality, epitaxial $SrTiO_3$ (STO) membranes. Using a binary oxide sacrificial layer of SrO, we show successful epitaxial growth, exfoliation, and transfer of STO membranes. Epitaxial and single-crystalline growth of STO on (Ba,Ca)O layer is also achieved. Direct growth of STO on (Ba,Ca)O layer resulted in polycrystalline films. This issue was addressed by inserting a thin diffusion barrier layer consisting of epitaxial $BaSnO_3$ (BSO) or $CaSnO_3$ (CSO). The as-grown STO films are exfoliated by dissolving $SrO/BaO/Ba_{1-x}Ca_xO$ in deionized water, followed by the transfer onto various host substrates. A rapid (≤ 5 minutes) dissolution of the sacrificial layer was achieved when immersed in deionized water. By fabricating metal-insulator-metal (MIM) capacitors of STO membranes on an Au-coated Si substrate, we demonstrate a bulk-like intrinsic dielectric constant in these membranes. Scanning near-field optical microscopy (SNOM) further demonstrated bulk-like phonon modes in membranes transferred on a Si substrate.

**Results and Discussion**

We use hybrid MBE to grow STO and the sacrificial layer of SrO, details of which are provided in the Materials and Methods section. Figures 2a and 2c show high-resolution x-ray diffraction (HRXRD) of 11 nm STO/5 nm SrO grown on a LaAlO$_3$ (LAO) (001) substrate and 120 nm STO/2 nm SrO grown on a LSAT (001) substrate, respectively. Phase-pure, epitaxial, single crystalline films with the out-of-plane lattice parameters ($a_{op}$) of 3.915 ± 0.002 for 11 nm STO on 5 nm SrO/LAO and 3.942 ± 0.002 Å for 120 nm STO on 2 nm SrO/LSAT were observed. The $a_{op}$ of 5 nm SrO film on LAO was 5.129 ± 0.002 Å, whereas no obvious peaks were seen for 2 nm SrO on LSAT. This can likely be explained owing to the thinner SrO layer on LSAT (001). Relative to the bulk lattice parameters ($a_{STO-bulk}$ = 3.905 Å and $a_{SrO-bulk}$ = 5.139 Å), both structures yielded partially strained STO films and a nearly fully relaxed SrO. The SrO layer grows with an in-plane rotation of 45° ($a_{SrO-bulk}/\sqrt{2}$ = 3.634 Å) to minimize the lattice mismatch. The $\phi$-scans in Figure 2e confirm the epitaxial relationship of (001)$_{STO}$[100]$_{STO}$//(001)$_{SrO}$[110]$_{SrO}$ and (001)$_{SrO}$[110]$_{SrO}$//(001)$_{LAO}$[100]$_{LAO}$ for STO on SrO and of SrO on LAO, respectively. The cross-sectional STEM images (Figure S2) further confirmed this observation in addition to revealing structural defects (threading dislocations), which we attribute to the strain relaxation and sample deterioration due to the hygroscopic nature of sacrificial layer.

The insets of Figures 2a and 2c show the streaky reflection high-energy electron diffraction (RHEED) patterns after SrO growth and STO growth on both substrates, revealing single-crystalline and epitaxial films with smooth surface morphology. The RHEED intensity as a function growth time (Figure S3), reveals a step-flow growth mode for STO on 5 nm SrO/LAO and a layer-by-layer growth mode for STO on 2 nm SrO/LSAT substrate. Atomic force microscopy (AFM) images of STO surface confirmed a smooth surface morphology (see the insets of Figures 2b and 2c). This observation is further consistent with the well-defined Kiessig fringes observed in the x-ray reflectivity (XRR) measurement (Figure 2b) of the same 11 nm STO/5 nm SrO/LAO (001) sample. Fitting of the XRR data (black solid line in Figure 2b) yielded a film thickness of 11 nm for STO and 5 nm for SrO, values which are consistent with the expected layer thicknesses. Note that all the measurements were conducted soon after the samples were taken out of the vacuum chamber to avoid deterioration of the SrO sacrificial layer when exposed to air.

To investigate this effect, we performed a detailed study using a thickness series of STO/SrO/LAO(001). Our goal is two-fold: (1) to examine the potential effect of SrO deterioration on STO crystallinity, and (2) to see whether or not an ex-situ rocking curve measurement be used to examine the quality of epitaxial growth of STO on a sacrificial SrO layer. The latter is particularly important as the rocking curve's full-width at half maximum (FWHM) can be broadened after growth due to the deterioration of the underlying SrO layer. We found exactly the same: Figures S4 and S5 show the results from the HRXRD of $t_{STO}$ nm STO/ 5 nm SrO/LAO (001), and 120 nm STO/ $t_{SrO}$ nm SrO/LAO (001), respectively with varying $t_{STO}$ and $t_{SrO}$. The analyses of data show the degradation of SrO can be substantially reduced when a thin SrO ($\leq$ 5 nm) layer is covered with a thick STO layer (~ 120 nm). For instance, the FWHM of a 120 nm STO continues to decrease, with reducing SrO layer thickness, reaching a value of 0.46° which is identical to the FWHM of the film grown without any SrO layer. For a detailed discussion, we refer the readers to the supplementary information (Figures S3-S4). Since LAO substrates are known to be twinned, we explored LSAT (001) to evaluate the crystallinity of our STO film grown on SrO/LSAT (001). Figure 2d shows representative rocking curves around the STO (002) peak for both 120 nm and 10 nm STO grown on a 2 nm SrO on LSAT (001) substrate. A substantially narrower FWHM of ~ 0.15° was measured for STO films. Unlike films grown on LAO substrate, even a thinner (10 nm) STO film on 2 nm SrO/LSAT (001) didn't show appreciable degradation. Future experiments should examine the effect of substrate on sacrificial layer degradation. Nonetheless, it is prudent to exercise caution when assessing the quality of epitaxial growth in cases where the underlying layer is sensitive to air, as one can deduce from these observations. To mitigate this concern, the practice of transferring samples within a glove box and conducting in-situ diffraction studies can also be adopted.

To further assess the adaptability of binary oxides as sacrificial layers, we chose to focus on (Ba,Ca)O due to its close lattice match with STO (001) and its high hygroscopic nature, which facilitates quicker dissolution in deionized water. Our initial attempts to directly grow STO on a 16 nm BaO/STO (001) substrate resulted in the formation of polycrystalline films (see Figure S6). We attribute this outcome to the undesirable intermixing between BaO and STO, creating a suboptimal template for epitaxial growth (refer to our discussion in Figure S1). To address this challenge, we successfully introduced a barrier layer of either BSO or CSO. The results of epitaxial, single crystalline STO growth using BSO are presented in Figure S7. Here, we will discuss the findings related to the utilization of CSO as a barrier layer.

Figure 2f shows HRXRD of an as-grown 120 nm STO/6 nm CSO/30 nm Ba$_{1-x}$Ca$_x$O/LSAT (001) heterostructure. These thicknesses were chosen arbitrarily. Details for growth are discussed in the Materials and Method section. CSO was chosen as a barrier layer because its pseudocubic lattice parameter (a$_{CSO-pc}$ = 3.944 Å) closely matches with STO. From HRXRD, the lattice parameter of STO is found to be 3.902 ± 0.002 Å (close to the bulk-value). Reciprocal space maps (RSM) can be used to determine the exact strain state; however, long scans were avoided to prevent deterioration of the (Ba,Ca)O layer upon air exposure. The RHEED images after each layer growth reveals epitaxial nature, and single crystallinity (Inset of Figure 2f). Given the hygroscopic nature of (Ba,Ca)O, we do not use rocking curve FWHM as a measure of film crystallinity. Conversely, we show in Figure S8, HRXRD of a 120 nm STO/6 nm CSO/30 nm $Ba_{0.984}Ca_{0.017}O$/LSAT (001) with and without a silicon capping layer (18 nm). Consistent with our prior discussion in Figures S4-S5, these results again show degradation of (Ba,Ca)O as evident in the broader FWHM.

Furthermore, we show in Figures 2g-2h our ability to systematically tune the lattice parameter of the sacrificial layer using a solid solution between BaO and CaO. The wide-angle X-ray diffraction scans for all alloyed samples with structure 120 nm STO/6 nm CSO/30 nm $Ba_{1-x}Ca_xO$/LSAT (001) are shown in Figure S9. The lattice parameter of $Ba_{1-x}Ca_xO$ is controlled by changing the beam equivalent pressure (BEP) of Ba to BEP of Ca ratios, resulting in different compositions and therefore, the out-of-plane lattice parameter. The latter is in excellent agreement with Vegard's law assuming bulk lattice parameters (Figure 2h).

Following the successful epitaxial growth of STO on three distinct sacrificial layers, namely SrO, BaO, and $Ba_{1-x}Ca_xO$, we developed a methodology for the exfoliation and transfer procedure, as outlined in the Materials and Methods section. Figure 3a provides a schematic representation of the exfoliation process. Employing this method, we achieved the transfer of millimeter-sized membranes. However, similar to the challenges encountered with complex sacrificial layers, our process also led to the formation of microcracks, wrinkles, and bubbles during the creation of large-area membranes measuring 5 mm × 5 mm. These issues are attributed to the exfoliation and transfer process [48–50]. However, our process routinely yielded an uncracked

STO membrane or STO/CSO membrane of 0.5 mm × 0.5 mm (see inset of Figure 3b), sometimes exceeding 1 mm × 2 mm (see inset of Figure S10). When compared to a 20-40 nm $Sr_3Al_2O_6$ sacrificial layer, which typically takes one day to dissolve [19], our findings demonstrate that 2-5 nm SrO can dissolve within a day, a 15 nm SrO layer can dissolve in less than 30 minutes, and a 60 nm SrO layer can dissolve in less than 10 minutes in water at 80°C (Figure 3d). Furthermore, a 16 nm BaO layer dissolves in deionized water within a few minutes (less than or equal to 5 minutes), and the 30 nm $Ba_{1-x}Ca_xO$ sacrificial layer (x = 1.7 – 9 at%) also dissolves within a few minutes (less than or equal to 5 minutes) when exposed to water at 80°C. This is in contrast to the long dissolution time of 10 hours for 60 nm BaO shown in the previous work [37]. Collectively, our results underscore the rapid membrane synthesis enabled by the faster dissolution kinetics of binary oxides.

HRXRD scans confirm the phase-pure, single-crystalline STO membrane of varying thicknesses on Au-coated Si substrate using SrO sacrificial layer, with an out-of-plane lattice parameter of 3.905 ± 0.002 Å, identical to that of the bulk-single crystal (Figure 3b). We also obtained identical results for STO membranes that were transferred onto dissimilar substrates such as plastic tape, Si and metal-coated Si wafers (Figure S10-11). It is intriguing to note that even 2 nm SrO layer was dissolved and resulted in a membrane with size 3 mm × 4 mm with uncracked region of size 0.25 mm × 0.25 mm (Figure S10b). Figure 3c shows HRXRD of 120 nm STO/6 nm CSO on Si. These films also yielded nearly bulk-like lattice parameters irrespective of their growth on $Ba_{1-x}Ca_xO$ sacrificial layer with varying x. A similar result was obtained from the 100 nm STO/ 2 nm BSO/16 nm BaO/STO (001) that was exfoliated by dissolving BaO layer, and transferred on Si (Figure S12). In a more detailed investigation of the atomic structure of these membranes, we successfully transferred 4-nm and 40-nm-thick STO membranes, produced with a 5 nm SrO layer, directly onto a TEM grid. Figures 3e-h present plan-view HAADF-STEM images, accompanied by energy dispersive x-ray spectroscopy (EDX), confirming the creation of phase-pure, single-crystalline STO membranes. To further assess the STO membrane's top surface and the interface with the SrO layer, we conducted AFM measurements. For the investigation of the interfacial STO/SrO region, we deposited amorphous Si on top of the STO, and then the membrane was transferred onto a thermal release tape, as illustrated in Figure 3i. Subsequently, we obtained an AFM image (with the rms roughness of 0.22 nm) of an uncracked region of the exfoliated STO membrane/amorphous Si/thermal release tape, representing the interface between the STO and SrO layer. The AFM analysis of the top surface revealed the rms roughness of 0.34 nm.

Finally, as a more sensitive measure of structural quality, we performed capacitance measurements as a function of frequency by fabricating a parallel-plate electrode geometry. We used photolithography to pattern a circular array of metallic electrodes followed by electron-beam evaporation to deposit Au electrodes. A device schematic along with a top-view optical image of the fabricated MIM device structure are shown in the insets of Figure 4b and 4c, respectively. We measured the impedance and the phase angle as a function of frequency at 300 K for a 120 nm STO membrane (Figure 4a). The complex impedance ($Z$) is a function of resistance ($R$) and reactance ($X_c$), such that $Z = R + iX_c$. For an ideal capacitor, $R = 0$ and $Z = X_c$, and the resulting current is out-of-phase (lagging) with the applied AC voltage with a phase angle of -90°. Figure 4b shows the impedance measurement for a 120-nm-thick STO membrane. Since the phase angle remains -90° for most of the frequency range ($f$), we can treat $Z = X_c$ and extract the capacitance ($C$) from the linear fit using the relationship shown in eq. 1.

$$\log(|X_c|) = -\log(2\pi f) - \log(C) \qquad [1]$$

Treating the structure as a parallel-plate capacitor, we can extract the static dielectric constant ($\varepsilon_r$) using eq. 2, where $t_{STO}$ is the membrane thickness, $A$ is the area of the top electrode, and $\varepsilon_o$ is the permittivity of free space.

$$\varepsilon_r = \frac{C\, t_{STO}}{\varepsilon_o A} \qquad [2]$$

Experimental data and the fitting results are shown in Figure 4b, and yield an effective dielectric constant of 40 for a 120-nm-thick thick STO membrane. This value is smaller than the dielectric constant of a bulk STO single crystal or an epitaxial STO films [51], and this result can be attributed to an interfacial effect between membrane and electrode. However, before we tackle the origin of the small dielectric constant, it is perhaps worth noting that our STO membranes are not post-annealed after growth and the exfoliation/transfer process, and the fact that the phase angle is -90° over a wide range of frequencies indicates a negligible leakage and, therefore, fewer oxygen vacancies. It may also be conceivable that there exists a gap (i.e. the transferred membrane is only weakly attached to the host substrate) at the interface between the transferred STO membrane and the metal electrode that affects the leakage currents [52]. To eliminate the role of an interfacial contribution to the measured dielectric constant of our STO membrane, we define the measured capacitance as a sum of the intrinsic capacitance ($C_{intrinsic}$) and the interfacial capacitance ($C_{interface}$), given by eq. 3. The interfacial capacitances come from the interfaces formed between the transferred STO and the bottom electrode ($C_{interface1}$) and between the transferred STO and the top electrode ($C_{interface2}$).

$$\frac{1}{C_{effective}} = \frac{1}{C_{intrinsic}} + \frac{1}{C_{interface}} \qquad [3]$$

The value of the intrinsic dielectric constant can be extracted by rearranging the above equation and plotting the measured capacitance ($A/C_{effective}$) as a function of membrane thickness ($t_{STO}$) as given by eq. 4.

$$\frac{A}{C_{effective}} = \frac{t_{STO}}{\varepsilon_{intrinsic}\, A} + \left(\frac{A}{C_{interface}}\right) \qquad [4]$$

To this end, we grew a series of STO membranes with thicknesses ranging from 10 nm to 422 nm and transferred them onto Au-coated Si substrates. MIM capacitor structures were fabricated, and their dielectric properties were characterized using an impedance analyzer (data shown in Figure S13). Figure 4c shows the experimental $A/C_{effective}$ as a function of $t_{STO}$ along with a linear fit. Using the model shown in eq. 4, we extract an intrinsic dielectric constant of 280 (similar to bulk value ~ 300). Whether the interfacial contribution comes from the weak interaction between the STO membrane and the metallic electrodes, interface reconstruction, or the air/vacuum gap remains unclear and should be investigated in future studies.

Infrared (IR) spectroscopy provides a powerful means to probe crystallographic order and symmetry in perovskites such as STO, which are quantified most simply through the frequency ($\omega$)-dependent complex-valued infrared permittivity $\varepsilon(\omega)$. To compare with the established permittivity of bulk STO (40), we examined $\varepsilon(\omega)$ of a thin STO membrane using nano-scale Fourier transform infrared spectroscopy (nanoFTIR). We targeted a 16-nm-thick single STO membrane and an artificially-created stack consisting of two 16-nm-thick STO membranes assembled at an *arbitrary* twist angle (Figure 4d), developed by reusing the substrate post membrane exfoliation (Figure S14). The stack was developed unintentionally as a result of the exfoliation and transfer process. Figures 4f-g present nanoFTIR spectroscopy of the STO membrane, revealing a clear phonon resonance corresponding to an optical phonon resonance. We associate this mode with the highest energy infrared-active transverse optical phonon of $F_{1u}$ symmetry (and often labeled TO4) that is well documented in bulk STO [53].

To extract infrared optical constants describing our quasi-2D STO membrane, we fit our nano-spectroscopy data using a quantitative model of near-field spectroscopy [54]. A 6-nm height differential between the topographic step-edge height measured by our concurrent AFM measurements and the nominal growth thickness (Figure 4d) was ascribed to a residual layer of polymethyl methacrylate (PMMA) following the transfer process. We included this PMMA layer in a multilayer model of the membrane atop silicon to simulate the expected frequency-dependent optical interaction with our probe [55]. Figure 4e shows the complex dielectric function extracted from the single-layer membrane, which supplies a simultaneously excellent fit for our measurements of the double-layered membrane as well. A negative real permittivity, emerging between the transverse optical ($\omega_{TO}$) and longitudinal optical ($\omega_{LO}$) phonon frequencies (the so-called Restrahlen band of TO4), enhances the probe-membrane optical interaction and, thus, the amplitude of scattered infrared light resolved spectroscopically in Figure 4f [56]. Meanwhile, for the phase shift of scattered light observed in Figure 4g, results from optical absorption of IR near-fields near the surface optical phonon energy ($\omega_{SO}$) where $\varepsilon \approx -1$ facilitates an infrared resonance (a so-called *surface phonon polariton*)[57]. This absorption is enhanced in the bilayer membrane owing to the doubled cross-sectional thickness for optical interaction. Figure 4d presents a nano-absorption map (rendered in false color) obtained by IR nano-imaging across the single- (rendered green in false color) and double- (red) layered membranes at their $\omega = 740$ cm$^{-1}$ phonon resonance, showcasing the sensitivity of near-field imaging and spectroscopy to membrane thickness. Meanwhile, the frequencies $\omega_{TO}$ and $\omega_{LO}$ extracted by our fit are identically consistent with those of unstrained bulk single-crystal STO, whereas the free parameter of our fit, the Ti-O phonon scattering rate ɣ, is found to be 1.9 greater than that identified from bulk single-crystals [58]. This indicates modest disorder in local crystallinity through the volume of our membranes relative to that reported in bulk crystals (40).

In summary, we demonstrate the potential of using binary alkaline earth oxides as a sacrificial layer for synthesizing perovskite oxide membranes. We employ hMBE to grow an epitaxial single-crystalline STO film on sacrificial layers consisting of crystalline SrO, BaO, and $Ba_{1-x}Ca_xO$ films. We demonstrate that STO growth on BaO results in polycrystallinity and can be avoided by coherently growing a thin barrier layer of CSO or BSO. These films are then transformed into millimeter-sized membranes by dissolving the SrO or BaO or $Ba_{1-x}Ca_xO$ sacrificial layer in deionized water, allowing for their integration with various structurally and chemically dissimilar substrates. Notably, the use of binary oxides as a sacrificial layer resulted in a rapid membrane synthesis (≤ 5 min) enabled by the faster dissolution kinetics of binary oxides. The intrinsic dielectric properties of the membranes are characterized using an impedance analyzer, and the phonon modes are observed using near-field spectroscopy. The results show that the dielectric properties and phonon modes are identical to those of bulk single-crystal STO, indicating that the membranes possess bulk-like structural and dielectric properties. The versatility of this approach is further demonstrated by successfully developing membranes of STO/CSO, STO/BSO, and STO using a family of binary oxide sacrificial layers.

**Materials and Methods**

**Growth and structural characterization**

We use a hybrid MBE system (Scienta Omicron Inc) to grow STO and SrO films on the LAO (001) substrate (MTI corporation). All films were grown on a 5 mm × 5 mm substrate. An effusion cell containing 99.99% Sr source was used to supply strontium. A liquid precursor of titanium tetra isopropoxide – TTIP (99.999% Sigma-Aldrich) is used to supply Ti that is injected inside the MBE system with the help of a line-of-sight gas injector (E-Science Inc) and a customized gas-inlet system controlled via a linear-leak valve and a Baratron capacitance monometer (MKS Instruments Inc). An RF oxygen plasma source was used for supplying oxygen. Oxygen plasma was operated at 250 W at an oxygen pressure of $5 \times 10^{-6}$ Torr. The beam equivalent pressure (BEP) of Sr was fixed at $1.2 \times 10^{-7}$ Torr for STO and SrO growth. TTIP was fixed at a Baratron pressure of 135 mTorr for an adsorption-controlled growth of STO films [27,59]. The substrate temperature was set to 900 ˚C for STO and SrO growth. The substrate was cleaned with oxygen plasma for 25 minutes prior to the growth. RHEED (Staib Instruments) was used for the in-situ characterization of film growth. HRXRD was performed using SmartLab XE X-ray diffractometer (Rigaku), and AFM was done using Nanoscope V Mulitmode 8 (Bruker). HAADF-STEM images and STEM-EDX elemental maps were obtained using a Thermo Fisher Talos F200X equipped with a Super-X EDX spectrometer. HAADF-STEM images and STEM-EDX maps were acquired at 200kV and a beam current of 20-80 pA. The camera length was set at 125 mm and a probe convergence angle of 10.5 mrad was used.

The growth of $Ba_{1-x}Ca_xO$ sacrificial layer was performed at a substrate temperature of 900 ˚C by supplying Ba and Ca using an effusion cell containing 99.99% Ba source and 99.99% Ca source, respectively, and oxygen using oxygen plasma operated at 250 W with oxygen pressure of $5 \times 10^{-6}$ Torr. The growth of CSO was performed using a Ca effusion cell fixed at a BEP of $9 \times 10^{-9}$ Torr and a hexamethylditin (HMDT) precursor operated at 250 mTorr Baratron pressure. A customized gas-inlet system similar to TTIP was used for HMDT precursor. Following CSO growth (at 900 ˚C), STO was grown with the same growth conditions as discussed earlier in this section.

**Exfoliation and Transfer**

**For SrO and BaO layer**

All samples were exfoliated within a day of being removed from the vacuum chamber. The first step is to apply support to the STO/$A$O/LAO sample ($A$ = Sr, Ba). The sample is spin-coated with a polymer supporting layer of polymethyl methacrylate (PMMA) at 3000 rpm for 60 seconds, and a thermal release tape with a square cut is applied on top of the PMMA. The sample was placed in the DI water for the duration of 24 hours that was heated to a temperature of 60 °C. Although higher temperatures (80 ˚C) helped dissolve a 15 nm thick SrO layer within 30 minutes, lower temperatures and high time were used to avoid potential damage to the exfoliated film from bubbling at a higher temperature. SrO dissolves via reaction $SrO + H_2O \rightarrow Sr(OH)_2$, the substrate falls down in the water or is gently lifted off, and a free-standing membrane of STO (with Tape/PMMA) is left floating in the container. The floating film (Tape/PMMA/STO) is subsequently scooped on top of the desired substrate. The PMMA supporting layer is removed using acetone, and thermal release tape is naturally detached. The same exfoliation and transfer process for SrO was used for all target substrates including Si and metal-coated Si.

**For $Ba_{1-x}Ca_xO$ layer**

Similar steps were followed for exfoliation and transfer of films grown on the $Ba_{1-x}Ca_xO$ sacrificial layer, except that PDMS or Tape was used as a supporting layer instead of PMMA. The samples were first attached with these supporting layers, followed by the dissolution of the sacrificial layer in water kept at 80 ˚C via reaction $Ba_{1-x}Ca_xO + (y)H_2O \rightarrow (1-x)Ba(OH)_2 + (x)Ca(OH)_2$. The substrate falls down in the water, resulting in free-standing STO/CSO heterostructures. The PDMS and Tape are removed from the membrane by heating the supporting layer/STO/CSO/host substrate stack to 120 ˚C.

**Fabrication**

Photolithography (lift-off process) was used to pattern an array of 100 µm diameter circles on the sample. First, the STO membrane on an Au-coated Si substrate was spin-coated with a photoresist (NR71 3000p) at 3000 rpm for 45 seconds. The sample was then baked at 120 °C for 1 minute. The photoresist, covered with a mask (having a circular pattern), was exposed to UV light for 20 seconds. After exposure, the sample was heated on a hot plate at 100 °C for a minute. The photoresist was removed by a developer (RD 6) creating a circular pattern. 400 nm thick Au electrodes were then deposited on the patterned sample using e-beam deposition. To avoid the delamination of the deposited Au on the STO membrane, ultra-sonication was done at low power for 1 minute in the resist remover (RR41). The photoresist was further removed gently by blowing the resist remover using a syringe on the pattern. The process was repeated 4 to 5 times slowly until the pattern was entirely developed while ensuring the Au electrode remained adhered to STO. Optical images were taken during and after the pattern development to ensure a sharp circular pattern was formed on membranes.

**Dielectric measurements**

An impedance analyzer (Keysight E4990A) connected to a probe station was used for impedance measurements. A standard 100 Ω resistor (Keysight E4990A-61001) was used for instrument calibration. The instrument was tested for open, short, and the standard 100 Ω before testing on the STO membrane.

**Nano-Infrared measurements**

STO membranes transferred to a Si substrate were investigated using a Park Systems NX-10 atomic force microscope (AFM) modified to function as a scattering-type near-field optical microscope (s-SNOM), using a home-built asymmetric Michelson interferometer with dual imaging and spectroscopy modalities. Mid-infrared light was sourced from a difference-frequency generation module pumped by dual optical parametric amplifiers of an ultrafast laser system (MIR, Alpha-HP, and Primus units from Stuttgart Instruments), supplying 500 picosecond infrared pulses with 50 cm$^{-1}$ bandwidth at 40 MHz repetition rate. This light was focused to the sharp tip of a metallic microcantilevered AFM probe, and subsequently scatters in a manner determined by the complex permittivity of the sample volume proximate to the tip, whose radius determines the optical spatial resolution. A cryogenically cooled mercury cadmium telluride photoconductive detector (Judson Teledyne) recorded scattered infrared light which we demodulated (HF2LI from Zurich Instruments) at the third harmonic of the AFM probe oscillation (at 75KHz, using an FM-PtSi probe from NanoSensors) for background suppression [60]. Placing the s-SNOM and sample at one arm of the interferometer allows simultaneously extracting the amplitude and phase of scattered light [61]. Nano-infrared imaging (Figure 4A) and spectroscopy measurements (Figure 4C-D) were conducted with time-resolved pseudo-heterodyne imaging (43) and nanoscale Fourier transform infrared spectroscopy methods [62], respectively. NanoFTIR spectra were synthesized by merging several distinct spectra acquired at incremental tunings of the characteristic energy of the MIR stage output (from 600 to 1000 cm$^{-1}$).

**Acknowledgments**

We thank Darrell Schlom and Sreejith Nair for the helpful discussion and comments. Synthesis of membrane and characterization (S.V. and B.J.) were supported by the U.S. Department of Energy through DE-SC0020211. Device characterization (fabrication and dielectric measurements) was supported as part of the Center for Programmable Energy Catalysis, an Energy Frontier Research Center funded by the U.S. Department of Energy, Office of Science, Basic Energy Sciences at the University of Minnesota, under Award No. DE-SC0023464. S.C. and Z.Y. acknowledge support from the Air Force Office of Scientific Research (AFOSR) through Grant Nos. FA9550-21-1-0025, FA9550-21-0460. Part of the work (J.W., S.J.K., and B.J.) was supported by the AFOSR through Grant FA9550-23-1-0247. Film growth was performed using instrumentation funded by AFOSR DURIP award FA9550-18-1-0294. J.S. and K.A.M. were supported partially by the UMN MRSEC program under Award No. DMR-2011401. Nano-infrared investigation of 2D materials was facilitated with support by AFOSR through Award No. FA9550-22-1-0330 under the DEPSCoR program. Parts of this work were carried out at the Characterization Facility, University of Minnesota, which receives partial support from the NSF through the MRSEC program under award DMR-2011401. Exfoliation of films and device fabrication was carried out at the Minnesota Nano Center, which is supported by the NSF through the National Nano Coordinated Infrastructure under award ECCS-2025124.

**Author Contributions:** S.V., S.C., and B.J. conceived the idea and designed the experiments. S.V. grew the films. S.V., S.C. developed the exfoliation and transfer process. S.V., S.C., and Z.Y. performed structural characterization and electrical testing. J.W. assisted with the dielectric measurements under the supervision of S.J.K. STEM/EDX measurements were performed by J.S. under the supervision of K.A.M. SNOM measurements and analysis of the optical data were done by L.T. under the supervision of A.M. S.V., and B.J. wrote the manuscript. All authors contributed to the discussion and manuscript preparation. B.J. directed the overall aspects of the project.

**Competing Interest Statement:** The authors declare no competing interests.

**Data and materials availability:** All data needed to evaluate the conclusions of the paper are present in the paper and/or the Supplementary Materials.

**Figures**

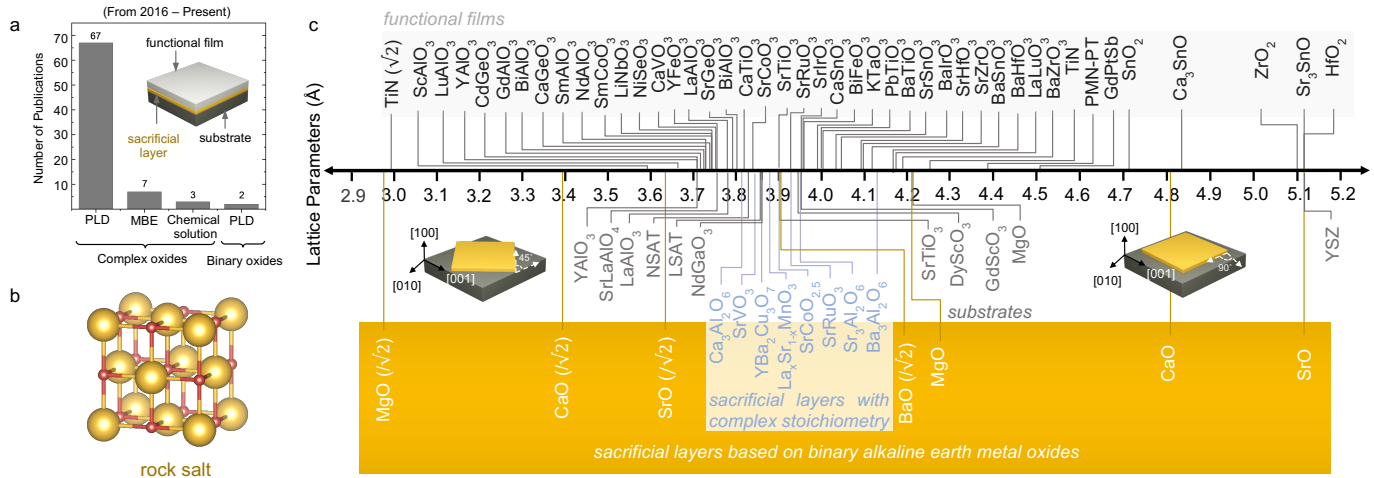

**Figure 1.** (a) Number of publications on epitaxial membrane synthesis using an epitaxial sacrificial layer from 2016 to present (August 2023). For complex sacrificial layers, there are 67 papers using PLD, 7 papers using MBE [10,31–36], and 3 papers using chemical solution deposition method [28–30]. There were only 2 papers using an epitaxial binary oxide using PLD. Inset shows a schematic illustrating the structure consisting of a functional film/sacrificial layer/substrate. (b) Binary alkaline earth metal oxide with a rock salt crystal structure. Alkaline earth elements are labeled in yellow, whereas oxygen atoms are labeled in red color. (c) Lattice parameters of sacrificial layers with complex stoichiometry (light blue), binary alkaline earth metal oxide sacrificial layers along with their growth orientations (yellow band), the commercially available substrates (grey), and several functional films (black). Pseudocubic lattice parameters are used for non-cubic perovskites and other functional materials. Lattice parameters are taken from [63–65] and are tabulated in Table S1 for readability.

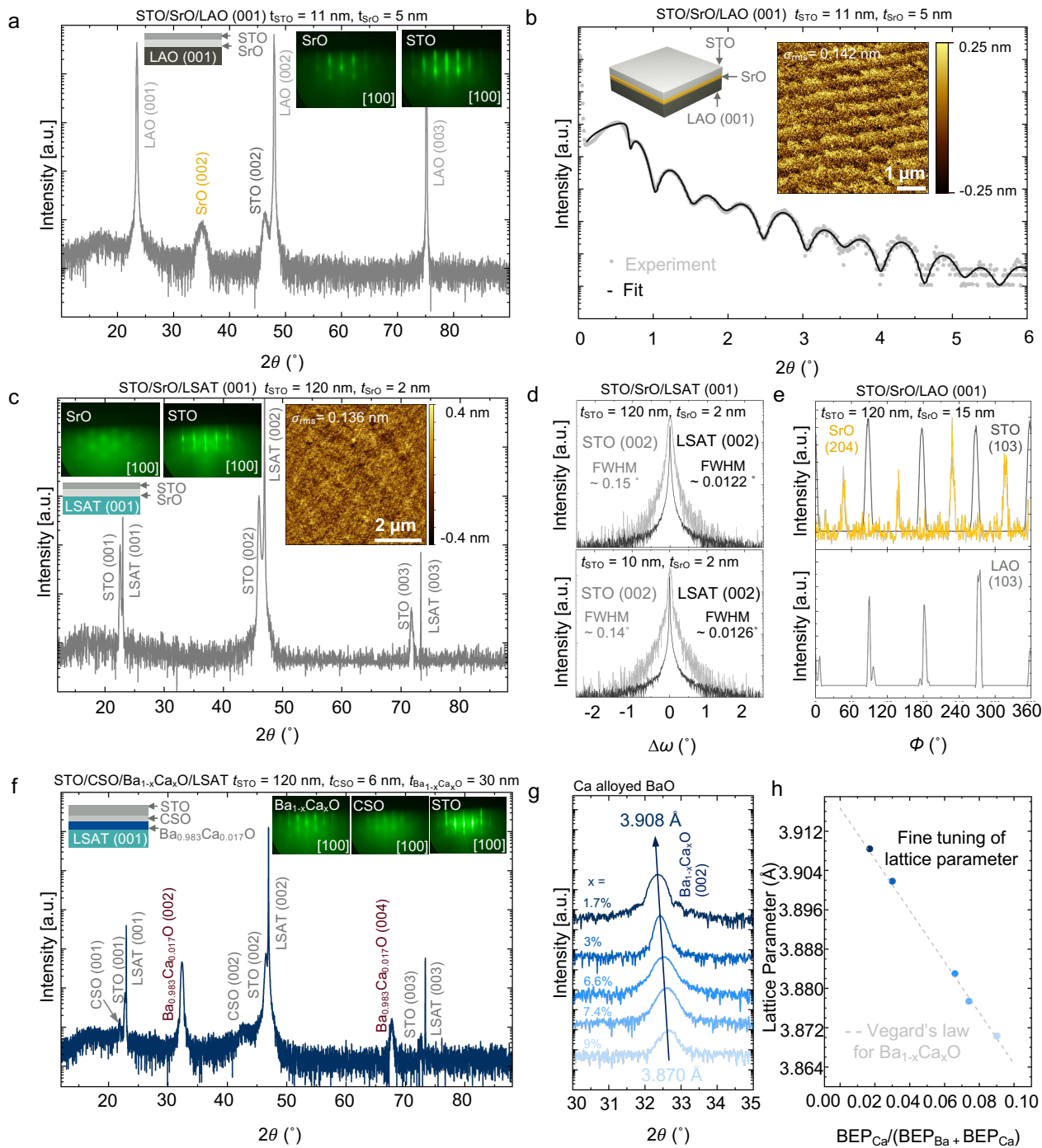

**Figure 2**. Structural characterization of STO films grown on alkaline earth metal oxide sacrificial layer. (a, c) Wide-angle X-ray diffraction 2θ-ω coupled scans for 11 nm STO/ 5 nm SrO/LAO (001) and 120 nm STO/2 nm SrO/LSAT (001) respectively. The inset shows a sample schematic

and RHEED images after STO and SrO growth along [100] directions. Inset of (c) shows an AFM image showing smooth surface morphology. (b) XRR showing Kiessig fringes. The insets show a sample schematic and AFM image of 11 nm STO/ 5 nm SrO grown on LAO (001) with atomic-step terraces. (d) rocking curves around STO (002) peak grown on 2 nm SrO/LSAT substrate. (e) $\phi$-scans around (103) peak of STO and LAO and (204) peak of SrO, showing an in-plane epitaxial relationship. (f) Wide-angle X-ray diffraction 2θ-ω coupled scan of 120 nm STO with 6 nm CSO buffer layer on $Ba_{0.983}Ca_{0.017}O$ sacrificial layer on LSAT (001) substrate. Inset shows a sample schematic and RHEED images after each layer growth (g) Fine-coupled X-ray diffraction 2θ-ω scan around $Ba_{1-x}Ca_xO$ (002) peak for a range of Ca at. % ranging between 1.7 to 9%. (h) The lattice parameter of $Ba_{1-x}Ca_xO$ extracted from HRXRD plotted as a function of normalized Ba BEP ratios. The dashed line represents Vegard's law for bulk $Ba_{1-x}Ca_xO$.

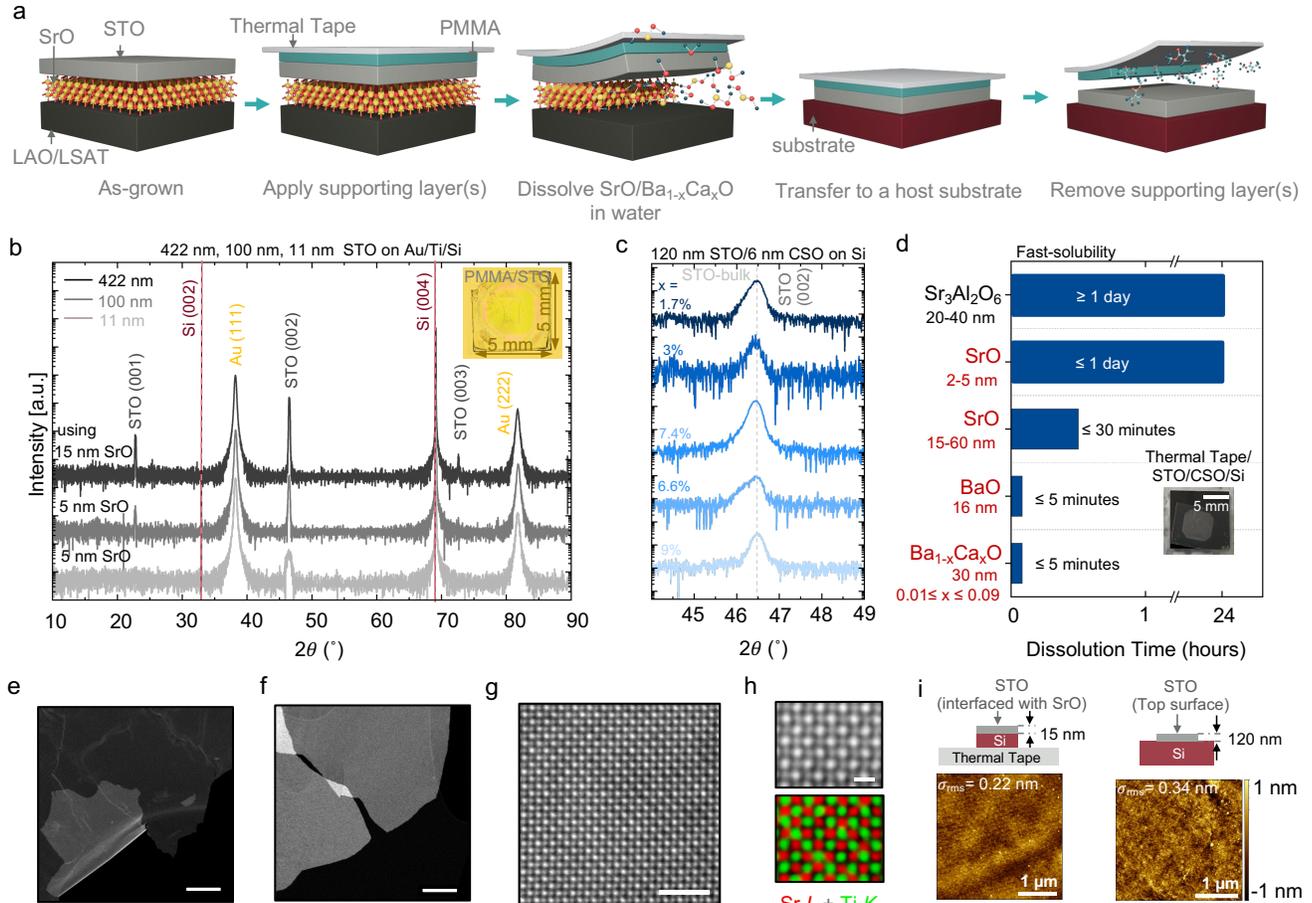

**Figure 3.** Exfoliation, transfer, and characterization of single-crystalline STO membranes. (a) Schematic illustration of the exfoliation and transfer process of STO membrane. (b) Wide-angle X-ray diffraction 2θ-ω coupled scan of a 422 nm, 100 nm, 11 nm thick STO membrane transferred on an Au-coated Si substrate using SrO sacrificial layer. Inset shows a representative optical image of an exfoliated PMMA/11 nm STO membrane transferred on Au/Ti/Si substrate (c) Fine-coupled X-ray diffraction 2θ-ω scan around STO (002) peak demonstrating the transfer of STO/CSO heterostructure membrane using $Ba_{1-x}Ca_xO$ sacrificial layer. (d) Plot of dissolution time of binary oxide sacrificial layer compared to complex sacrificial layer demonstrating faster membrane synthesis. Inset shows a STO/CSO membrane on Si. (e, f) HAADF-STEM images of 4 nm and 40 nm thick STO membranes transferred on a TEM support grid. The scale bars are 2 μm. (g) Atomic resolution plan-view HAADF-STEM image of a 40 nm-thick STO membrane. The scale bar is 2

nm. (h) A set of HAADF images and a complimentary EDX elemental maps. The scale bar is 5 Å. EDX maps are filtered for noise reduction and better visualization. (i) AFM images of STO membranes as indicated in the schematics after the exfoliation and transfer process.

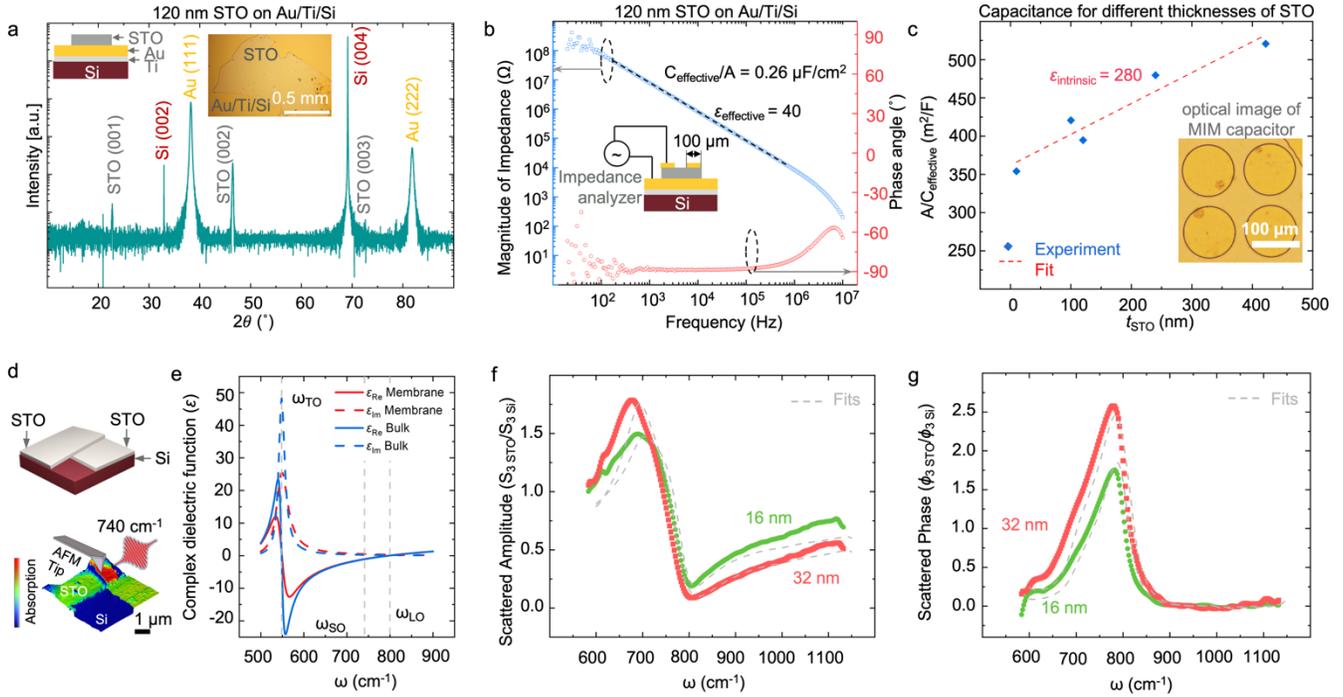

**Figure 4.** Dielectric properties of STO membranes. (a) Wide-angle X-ray diffraction 2θ-ω coupled scan of a 120 nm thick STO membrane transferred on an Au-coated Si substrate. Inset shows the sample schematic and a representative optical image demonstrating a millimeter-sized uncracked membrane. (b) Impedance spectroscopy measurement on a 120 nm STO membrane showing the magnitude of impedance and phase-angle as a function of frequency. Inset shows the a metal-insulator-metal (MIM) structure consisting of Au/STO/Au/Ti/Si(001). (c) A plot of $A/C_{\text{effective}}$ versus $t_{\text{STO}}$ for extracting the intrinsic dielectric constant. Inset shows a representative top-view of an optical image of the microfabricated MIM capacitor structure. (d-g) Near-field imaging and spectroscopy of STO membranes. (d) Schematic of the stacked bilayer of 16 nm membranes and accompanying nano-infrared phase contrast obtained by nano-imaging at 740 cm$^{-1}$ overlaid on simultaneously recorded surface topography. Higher phase shifts of scattered light (red in false color) indicate increased optical absorption. (e) Complex-valued permittivity for the STO membrane (dashed curves) obtained by fitting nanoFTIR spectroscopic data, as compared with that of bulk STO (solid curves). Characteristic energies $\omega_{\text{LO}}$, $\omega_{\text{TO}}$, and $\omega_{\text{SO}}$ are identical to those of bulk STO, whereas the scattering rate is comparably higher in our membranes. (f, g) Energy ($\omega$)-dependent amplitude ($s_3$) and phase ($\phi_3$) of probe-scattered light recorded at the third harmonic of probe oscillation by nanoFTIR spectroscopy; spectra are normalized to the uniform response of the adjacent silicon substrate (Si).